\documentclass[reprint,aps,prab]{revtex4-2}
\usepackage[english]{babel}
\usepackage[utf8]{inputenc}
\usepackage{amsthm}
\usepackage{mathtools}
\usepackage{xcolor}
\usepackage{graphicx}
\usepackage[T1]{fontenc}
\usepackage[pdftex, pdftitle={Article}, pdfauthor={Author}]{hyperref}
\DeclareMathOperator\erf{erf}

\begin{document}
\title{Creation and direct laser acceleration of positrons in a single stage}

\author{Bertrand Martinez}
    \email[]{bertrand.martinez@tecnico.ulisboa.pt}
    \affiliation{Golp/Instituto de Plasma e Fus\~{a}o Nuclear, Instituto Superior Técnico, Universidade de Lisboa, 1049-001 Lisbon, Portugal}

\author{Bernardo Barbosa}
    \affiliation{Golp/Instituto de Plasma e Fus\~{a}o Nuclear, Instituto Superior Técnico, Universidade de Lisboa, 1049-001 Lisbon, Portugal}

\author{Marija Vranic}
    \affiliation{Golp/Instituto de Plasma e Fus\~{a}o Nuclear, Instituto Superior Técnico, Universidade de Lisboa, 1049-001 Lisbon, Portugal}

\begin{abstract}
Relativistic positron beams are required for fundamental research in nonlinear strong field QED, plasma physics, and laboratory astrophysics. Positrons are difficult to create and manipulate due to their short lifetime, and their energy gain is limited by the accelerator size in conventional facilities. Alternative compact accelerator concepts in plasmas are becoming more and more mature for electrons, but positron generation and acceleration remain an outstanding challenge. Here we propose a new setup where we can generate, inject and accelerate them in a single stage during the propagation of an intense laser in a plasma channel. The positrons are created from a laser-electron collision at 90 degrees, where the injection and guiding are made possible by an 800 nC electron beam loading which reverses the sign of the background electrostatic field. We obtain a 17 fC positron beam, with GeV-level central energy within 0.5 mm of plasma.
\end{abstract}
\maketitle

\section{Introduction}

Plasma-based compact particle sources hold great potential for future applications.
While conventional accelerators are limited to a maximum acceleration gradient of $10 \, \rm MV/m$ due to the material damage threshold, using a plasma allows sustaining stronger fields ($ 10 \, \rm GV/m$), enabling short acceleration distances.
In addition to the high damage threshold, plasmas self-generate fields that are suited to focus electrons~\cite{PRLTajima1979}. 
Combining the acceleration and focusing forces, plasmas naturally provide both the acceleration and beam transport for the accelerated electrons. 
It is, however, arduous to extend this to positrons, as 
the accelerating structure developed for electrons is usually defocusing for a positron beam and can lead to beam breakup.
This makes the acceleration of positrons a particularly challenging problem.
Since the first proof of principle of positron guiding in a plasma~\cite{PRLNg2001,PRLHogan2003,PRLBlue2003,PRLMuggli2008}, only a limited number of experiments on positron acceleration followed~\cite{NatCorde2015,NCGessner2016,SRDoche2017,PRLLindstrom2018}.
Theoretical works proposed ideas to accelerate positrons by tailoring the driver~\cite{PRLJain2015,PRLVieira2014,PPCFLi2019,PRRHue2021} or by tuning the plasma profile~\cite{PoPChiou1995,PRLSchroeder1999,PRLWang2008,PRSTABKimura2011,SRYi2014,PRABDiederichs2019,PRLSilva2021,PREReichwein2022}.
This post-acceleration of positrons in a plasma enables to create high charge and high quality beams.
However, it is not a compact solution as it requires to create and transport a positron beam in a kilometre-long accelerator.

The upcoming 10-200 PW~\cite{EPJDHeinzl2009,CLEOPapadopoulos2019,OEYoon2019,RLEShaykin2014,HPLSERadier2022,HPLSEBromage2019,BookMourou2011,OLShao2020,EPJSTBashinov2014,OCMourou2012} laser infrastructures will bring us closer to answering the question "What is the best way to accelerate positrons in a plasma?".
The scattering of a relativistic electron beam with a multi-PW laser pulse is predicted to generate gamma rays and electron-positron pairs (through nonlinear Compton scattering and Breit Wheeler pair creation ~\cite{PRLBurke1997}).
Numerical studies have explored gamma-ray emission ~\cite{PoderPRX2018,PRLNeitz2013,PRLBlackburn2014,PRENiel2018} or positron creation~\cite{PRLSokolov2010,PRABLobet2017,PRABlackburn2017,arXivStreeter2022}, but still miss an in-depth investigation of positron acceleration.
To date, very few schemes have been proposed for positron creation and acceleration~\cite{PRLChen2010, PPCFYan2017,SRVranic2018,CPXu2020,CPHe2021} and they all rely on qualitative 2D and/or staged modelling.
Consistent modeling of all important steps (creation, injection, and acceleration) in 3D geometry is still missing.

\begin{figure}
    \centering
    \includegraphics[width=0.48\textwidth]{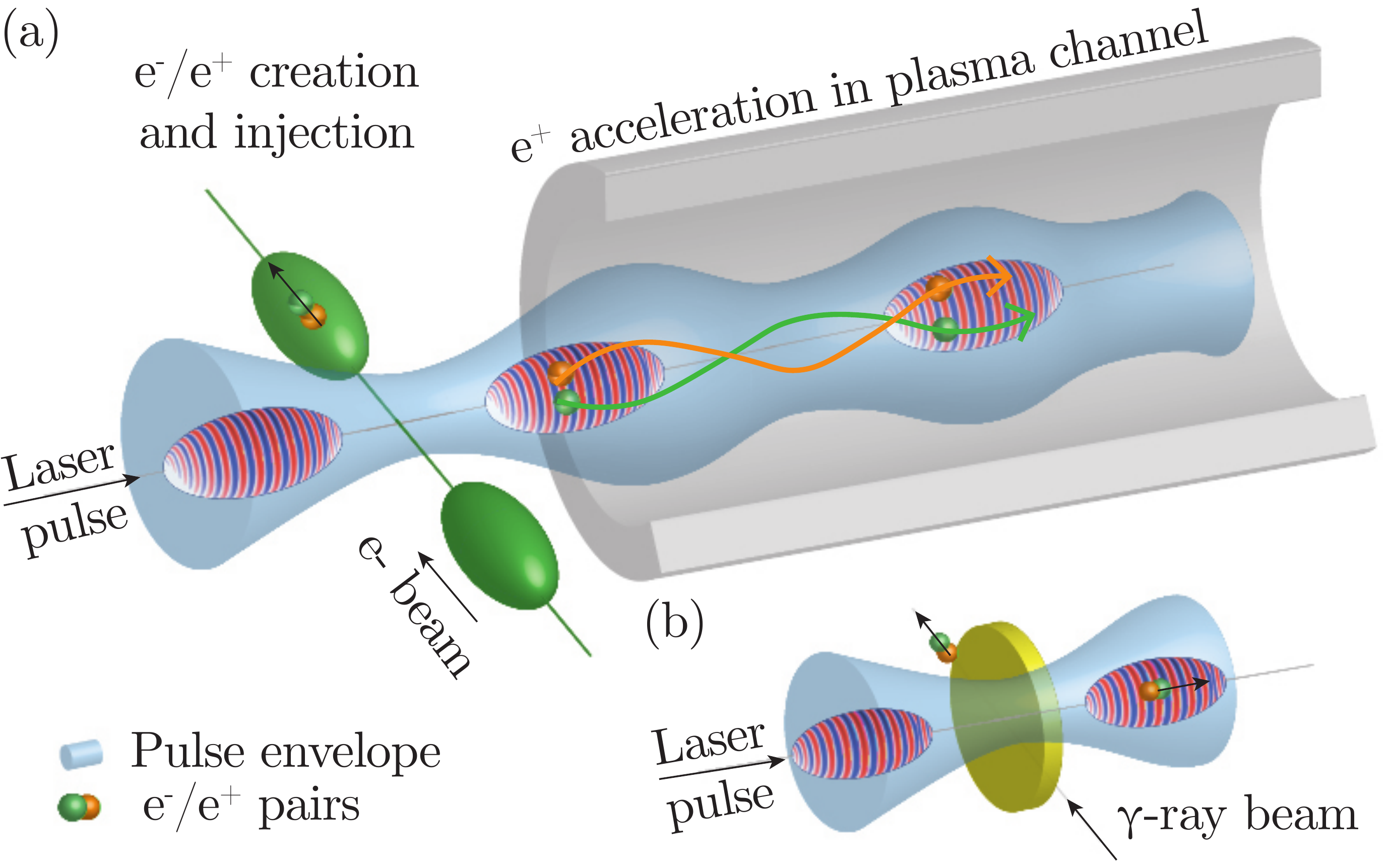}
    \caption{
    (a) Setup: electron-positron pairs (in green / orange) are created during the interaction of an intense laser pulse (blue/red) with a relativistic electron beam (green) at a $90^\circ$ incidence angle.
    A fraction of the pairs is injected and propagates with the pulse through the plasma channel, experiencing direct laser acceleration.
    (b) Quasi-3D modeling of this setup. A slab of $\gamma$-ray photons from electron-laser collision is initialized at $t=0$.
    }
    \label{fig:figure_1}
\end{figure}

In this work, we propose a new setup to obtain relativistic positron beams using next-generation laser systems (see Fig.~\ref{fig:figure_1}).
Positrons are created and accelerated during the interaction of an intense laser pulse ($\sim 5\times 10^{23} \, \rm W/cm^2$, 80 PW and 12 kJ of energy) with a relativistic electron beam at $90^\circ$ of incidence, as first introduced in Ref.~\cite{SRVranic2018}.
Even though most pairs are generated in the direction of the electron momentum, a fraction can be deflected by the laser field towards its propagation direction.
These positrons are then accelerated in vacuum, but their maximum energy is limited by the laser defocusing~\cite{SRVranic2018}.
Here, we derive a semi-analytical model for the number of positrons created, and show a few percent are deflected and available for acceleration. 
We propose to extend the acceleration distance of the pairs using a plasma channel, where the laser can be self-guided and the pairs can experience Direct Laser Acceleration (DLA)~\cite{PoPPukhov1999, PRLArevief2012, PoPArefiev2016, PoPKhudik2016, PREGong2020, NJPJirka2020, PREWang2021, PRABLi2021, NJPYeh2021}.
We demonstrate that for intense lasers propagating in a plasma channel, a dense central electron beam ($\sim 800~ \mathrm{nC}$) is self-injected and co-propagates with the laser on-axis ~\cite{PRLJi2014,PPCFVranic2018,PPCFWang2019}. This beam attracts positrons towards the channel center, enabling their acceleration to GeV energies.
We present an analysis of the energy gain induced by different field components, and show that a collimated ($ 50 \, \rm mrad $) positron beam of $17 \, \rm fC$ can be obtained within $400 \, \rm \mu m$ of laser propagation.
This work represents a proof-of-concept that positrons can be created, injected and accelerated over 0.5 mm of plasma.
Higher charge, energy and positron beam quality are expected by optimising initial conditions, and it will be a subject of a future study. 

\section{Positron guiding in a plasma channel}

The setup in Fig.~\ref{fig:figure_1} can be modeled using a Particle-In-Cell (PIC) code coupled with a Monte-Carlo module that accounts for pair creation and hard photon emission. 
A reliable description of laser guiding in a plasma channel requires modeling electromagnetic fields in 3D geometry.
However, this is out of reach of modern supercomputers as we estimated it would require 100 million cpu.hours and a total memory of 10 petabytes.
Quasi-3D modeling with the \textsc{osiris}~\cite{IPFonseca2002} framework is possible: an approach where the fields are represented in cylindrical coordinates $(z,r,\phi)$ using a Fourier decomposition in angular modes ~\cite{JCPLifschitz2009, JCPDavidson2015}.
The first two modes account for the axisymmetric self-generated channel fields (mode $m=0$) and for the non-axisymmetric linearly polarized laser field ($m=1$).
With this geometry, we cannot represent the electron beam at 90 degrees, but we can represent the photons it radiates. The assumptions we made and all simulation parameters are in the Supplementary Material~\cite{SupMat}.

\begin{figure}
    \centering
    \includegraphics[width=0.48\textwidth]{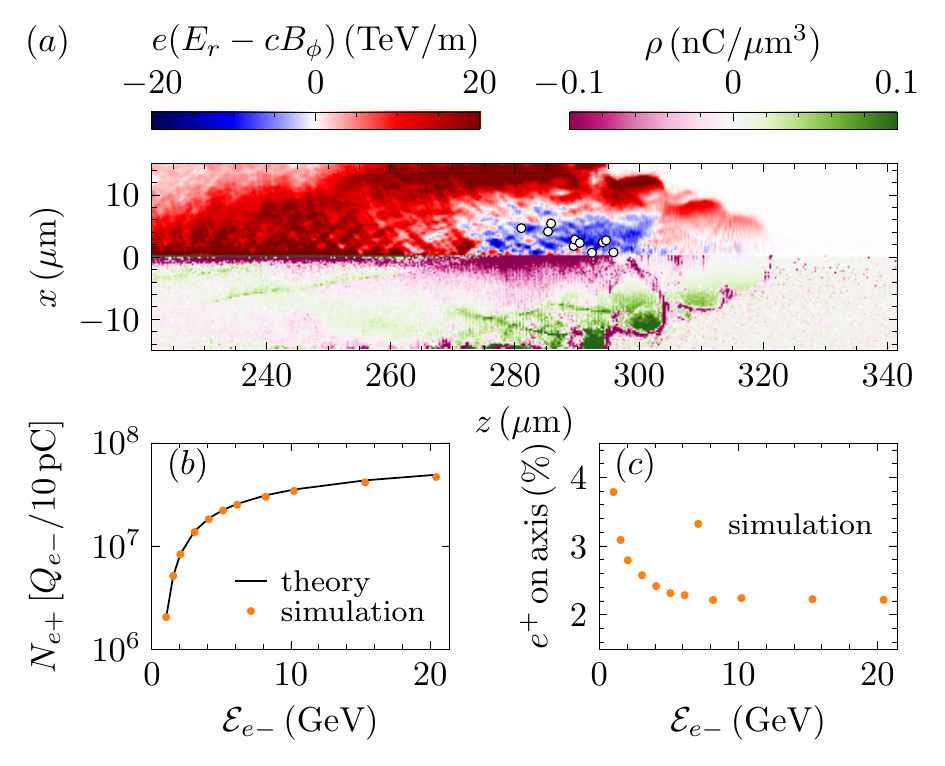}
    \caption{ Positron guiding and injection. (a) Electron beam loading after $280 \, \rm \mu m$ of propagation. Upper panel: channel fields contribution to the transverse Lorentz force on the positrons ($m=0$). Lower panel: net plasma charge density.
    (b) Number of generated positrons; (c) The fraction of positrons injected in the laser propagation direction as a function of the electron beam energy for a laser field amplitude $5\times 10^{23} \, \rm W/cm^2$ (circles).
    Panel (b) shows the number of positrons generated using an electron beam of 10 pC, and can be directly scaled to different beams (e.g. $Q_{e-} = 100 \, \rm{pC}$ would give 10 times more positrons).
    }
    \label{fig:figure_2}
\end{figure}

We consider a laser duration of $150 \, \rm fs$, peak intensity of $5\times 10^{23} \, \rm W/cm^2$ and a plasma channel with a radial density profile of parabolic shape (density $\sim 10^{19}$-$10^{21} \, \rm / cm^3$).
As the laser propagates in the channel, its ponderomotive force repels electrons toward the channel walls, forming a positive radial electric field $\mathbf{E}_c = +|E_c| \mathbf{r}$.
This electric field remains positive and persists as long as ion motion is negligible~\cite{PPCFWang2019}.
In addition, the laser electric field extracts electrons in its polarization direction from the channel walls.
The magnetic component of the laser can rotate these electrons along the channel axis enabling their injection and making them available for direct laser acceleration.
The tenuous electron beam formed drives a negative azimuthal magnetic field $\mathbf{B}_c = -|B_c| \boldsymbol \phi$.
The resulting transverse Lorentz force $|E_c|+|B_c|$ pulls the co-propagating electrons towards the axis.
The same force pulls positrons in the opposite direction, acting against their guiding.

In the context of DLA driven by a high-intensity laser pulse ($\geq 10^{23} \, \rm W/cm^2$), the central electron beam density increases with the propagation distance.
Ions can be injected and accelerated along the channel axis~\cite{PRLJi2014,NJPGelfer2021}, thus reducing the positive radial electric field $|E_c|$~\cite{PPCFVranic2018,PPCFWang2019}.
In addition, the electron beam loading is enhanced by a radiative trapping~\cite{PRLJi2014,PPCFVranic2018}.
An excess of negative charges is thus progressively built up at the channel center, inducing a negative charge separation field $\mathbf{E}_{e-} = -|E_{e-}| \mathbf{r}$, which can overcome the positive, self-generated field $\mathbf{E}_c$ formed by the initial expulsion of the electrons $|E_{e-}| > |E_c|$.
This superposition results in a radial channel electric field amplitude $ |E_c|-|E_{e-}| <0 $ guiding positive charges on axis.

We demonstrate the existence of a region focusing positrons using PIC simulations.
In the upper panel of Fig.~\ref{fig:figure_2}(a), we represent the transverse Lorentz force induced by the channel fields (mode 0) on positrons (white circles).
After $ 280 \, \rm \mu m$ of propagation, the force is negative in the blue region ($275\leq z \leq 300 \, \rm \mu m$ and $r\leq 10 \, \rm \mu m$).
The lower panel of Fig.~\ref{fig:figure_2}(a), displays the total charge density (including all plasma species)
where we observe an excess of negative charges associated with electron beam loading ($\simeq 800 \, \rm nC$).
The spatial structure of this beam generates a region of space where positrons are stably guided over 100s of micrometers.

\section{Positron creation and direct laser acceleration}

Now that we established the feasibility of positron guiding, we focus the discussion on positron creation, injection, and acceleration.
The number of positrons generated at the laser focus can be evaluated adapting a semi-analytical model~\cite{PRABlackburn2017,NJPAmaro2021,NJPMercuri2021} for our specific interaction geometry.
In Ref~\cite{PRABlackburn2017}, the authors derive the probability of a single photon decay $P_\pm$ in a head-on collision with a plane wave with a temporal envelope.
In the Supplementary Material~\cite{SupMat}, we generalize this model to an arbitrary angle of incidence.
We further enriched this model to account for a focused pulse interacting with a photon beam, which has a Gaussian shape in the z direction ($3 \, \rm \mu m$ FHWM) and is uniform in the radial direction.
Each photon in the beam experiences an effective peak field amplitude denoted $a$, which is smaller than the maximum pulse amplitude~\cite{NJPAmaro2021}. The photons can be binned according to the maximum field they interact with, which defines a new distribution $dN/da$. 
The total number of positrons is thus obtained by integrating the probability $P_\pm$ over the distribution $dN/da$.
Assuming that the interacting photons have a Quantum synchrotron spectrum $dN_{\mathrm{IC}}/d\gamma_\gamma dt$ of a monoenergetic electron beam with energy $\mathcal{E}_{e-}$ 
~\cite{RMPErber1966}, we get
\begin{equation} \label{equation:1}
    N_{e+} = \int_{\gamma_\gamma} \int_{a} \, \left( \frac{dN_{\mathrm{IC}}}{d\gamma_\gamma dt} \times \frac{dN}{da} \times P_\pm \right) \bigg/ \left( \frac{dN_{\mathrm{IC}}}{dt} \right)
\end{equation}
where $P_\pm$ is our generalised theoretical decay probability for one gamma ray in a plane wave~\cite{SupMat}.

The semi-analytical estimate in Eq.~\eqref{equation:1} provides the number of positrons that one could create within a plasma channel.
The accuracy for different electron beam energies is confirmed via PIC simulations displayed in Fig.~\ref{fig:figure_2}(b).
The peak laser amplitude is $5\times 10^{23} \, \rm W/cm^2$.
We considered electron beam energies $\mathcal{E}_{e-}$ between 1 and 20 GeV to cover the range available either from plasma or conventional sources.
The theory is valid in the limit where most pairs are created as a result of the decay of initial photons (no secondary pairs are accounted for).
As expected, Fig.~\ref{fig:figure_2}(b) shows that the number of positrons is an increasing function of the electron beam energy.
The number of pairs in the simulation is well-approximated by Eq.\eqref{equation:1}, which can therefore be used to quickly identify optimal incident electron beam energies for future experiments.
Our modelling indicates that the number of positrons (created and deflected) linearly depends on the charge of electrons interacting with the laser focal volume.
For instance, using a 100 pC laser-driven beam would generate 10 times more positrons than obtained in Fig.~\ref{fig:figure_2}(b).

\begin{figure}
    \centering
    \includegraphics[width=0.48\textwidth]{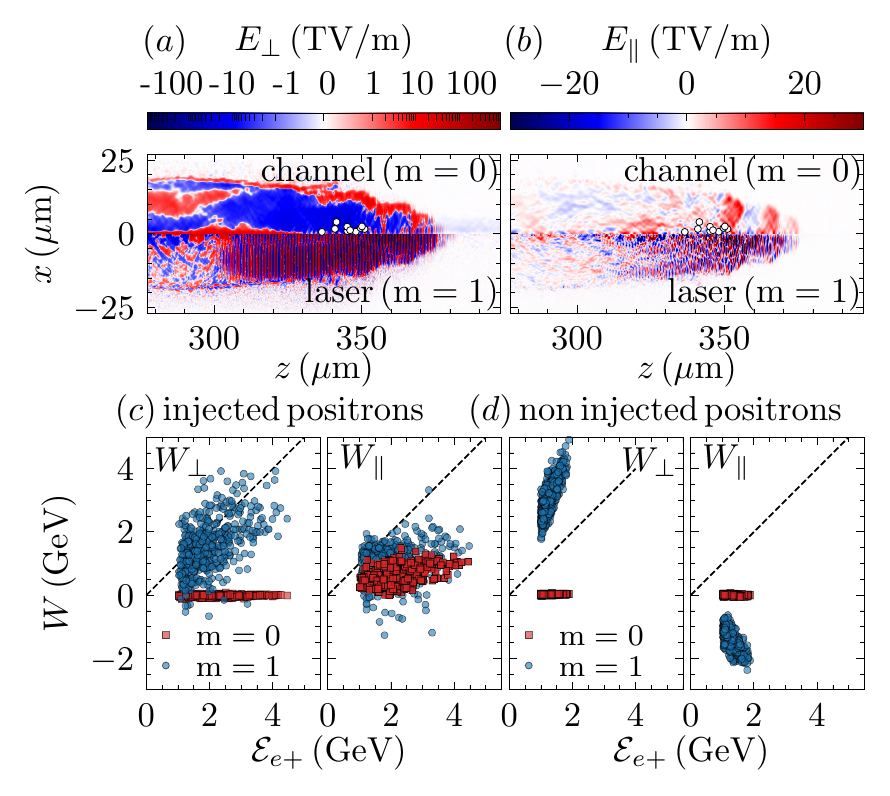}
    \caption{ Positron direct laser acceleration. 
    (a) Transverse electric field ($x$-direction). White circles represent a sample of injected positrons. Mode $m=1$ is associated with the laser component, while $m=0$ represents the channel field.
    (b) Longitudinal electric field ($z$-direction). 
    (c) Energy gain of injected positrons in transverse ($W_\perp$) and longitudinal directions ($W_\parallel$).
    (d) Energy gain of non-injected positrons.
    The energy gains $W$ represent the accumulated work in the channel fields ($m=0$, red squares) and the laser fields ($m=1$, blue circles) after a propagation distance of $340 \, \rm \mu m$.
    }
    \label{fig:figure_3}
\end{figure}

Following their creation, positrons have a momentum perpendicular to the laser propagation direction.
Simulations show that only a fraction of these can be deflected towards the laser propagation direction and later accelerated, see Fig.~\ref{fig:figure_2}(c).
These PIC simulations account for secondary pair production.
The efficiency of deflection is maximized ($4\%$) for the lowest value of incident electron beam energy $\mathcal{E}_{e-}\sim 1\, \rm GeV$.
Such beams produce on average lower energy positrons, which are more prone to be redirected in the laser propagation direction~\cite{SRVranic2018}.
This trend is confirmed in the Supplementary Material~\cite{SupMat}, where we derived an estimate of the number of low-energy positrons created.

Once injected within the channel, positrons can be affected by the laser field and the nonlinear plasma response.
Both laser and channel fields have a longitudinal and transverse electric component (with respect to the laser axis), shown in Fig.~\ref{fig:figure_3} (a-b).
We note that due to electron beam loading, the transverse electric field generated in the channel has a negative sign in a large region, attracting positrons on axis ($\sim 10 \, \rm TV/m$).
As it propagates, the laser pulse undergoes self-focusing~\cite{PFSun1987}. Depending on the background plasma density, the self-focusing dynamics can be affected by parametric plasma instabilities. In addition, our study is in the relativistic regime, where also radiation reaction effects can become important due to the high laser intensity. Even though the nonlinear interplay of all these effects is not possible to predict analytically, they are all naturally incorporated in the QED-PIC simulations, which we can use to model the laser pulse guiding.
We observed that the laser field amplitude is still strong ($ \gtrsim 200 \, \rm TV/m $) even after $340 \, \rm \mu m$ of propagation due to relativistic self-focusing.
Apart from strong transverse fields, both the longitudinal channel field (due to local charge separations) and the longitudinal laser field (due to relativistic self-focusing) are significant.

The total contribution of each field component for the positron energy gain ($W$) is studied in Fig.~\ref{fig:figure_3}(c).
We show a scatter plot for 500 injected positrons, where the x-axis denotes their final energy ($\mathcal{E}_{e+}$) and the y-axis shows the total work of different E field components.  
We consider positrons injected if they remain inside the plasma channel after $420 \, \rm \mu m$ of laser propagation.
The energy gains $W$ are decomposed in the perpendicular/longitudinal direction $W_{\perp/\parallel}=\int \mathbf{v}\cdot \mathbf{E}_{\perp/\parallel} \, dt$.
They are further split between the contribution of the channel fields ($m=0$) and the laser ($m=1$).
This analysis is repeated in Fig.~\ref{fig:figure_3}(d) for 500 non-injected positrons.
The first result is that injected positrons in Fig.~\ref{fig:figure_3}(c) achieve energies of $4 \, \rm GeV$, whereas the non-injected positrons in Fig.~\ref{fig:figure_3}(d) are limited below $2 \, \rm GeV$.
By definition, non-injected positrons are not trapped in the guiding structure in Fig.~\ref{fig:figure_2}(a). They experience a partial acceleration in the laser pulse, but they cannot be collected on a detector after the plasma channel because they leave the interaction region sideways.
Overall, the direct contribution of transverse channel fields for acceleration is small (though it is responsible for guiding).
The parallel component of the laser plays a role for all positrons, but the work performed is purely negative for positrons deflected outside the channel. 
For injected positrons in Fig.~\ref{fig:figure_3}(c) we observe that the energy gain from the transverse laser field component $W_\perp$ can be as high as $4 \, \rm GeV$, while the longitudinal energy gains both from the laser and the channel fields is limited below $1.5 \, \rm GeV$.
This indicates that DLA prevails over other possible mechanisms (\emph{e.g.} laser or beam-driven wakefield acceleration).
We also observed the forking structure characterising DLA~\cite{PPCFShaw2018,NJPHussein2021}. The two beam sections are ejected at an angle $\pm 10^\circ$ with a divergence of $50 \, \rm mrad$.

\begin{figure}
    \centering
    \includegraphics[width=0.48\textwidth]{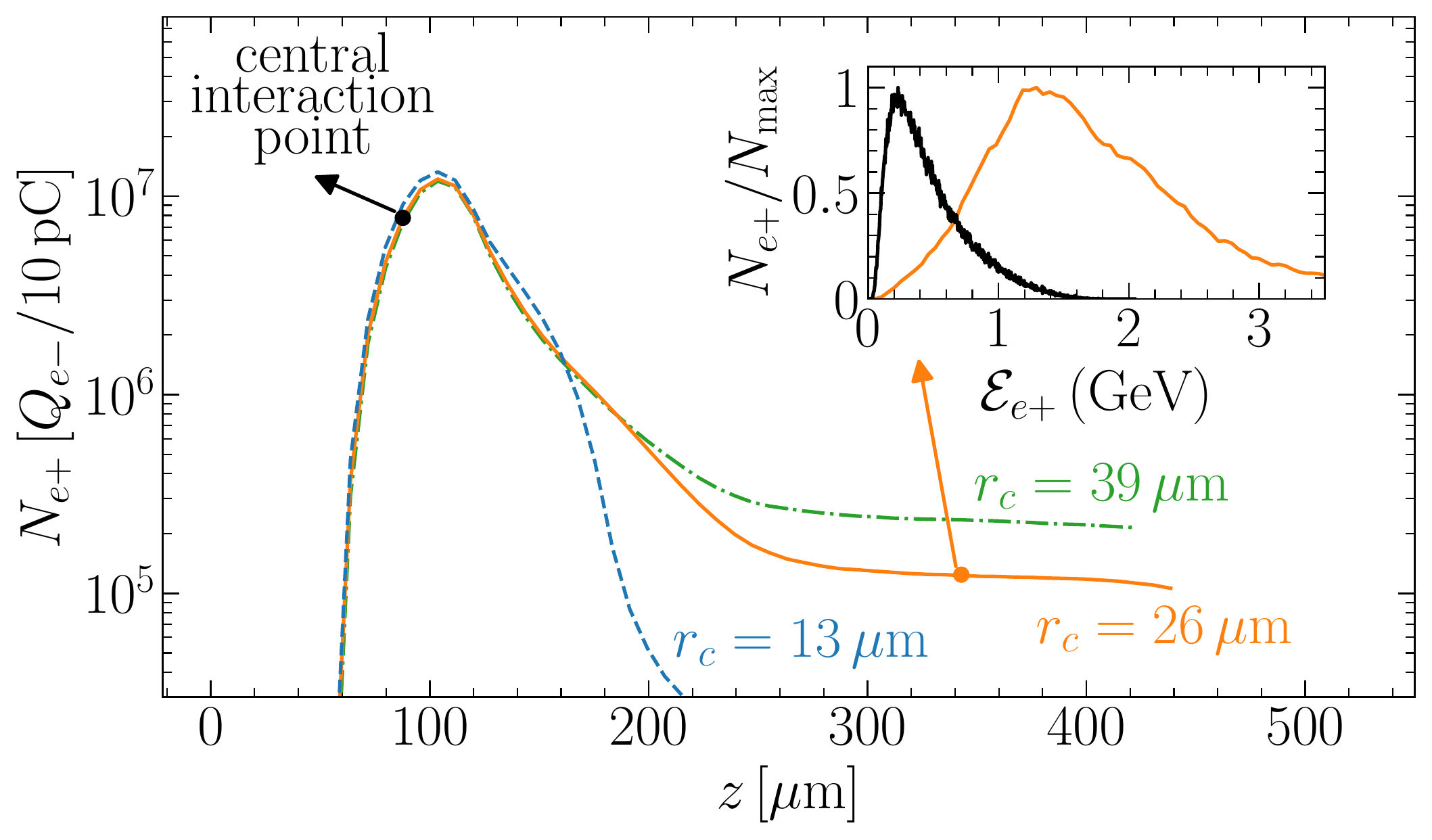}
    \caption{
    The number of injected positrons as a function of the laser propagation distance for different channel sizes.
    The black circle marks the central interaction point of the laser pulse with the photon beam.
    Inset: Positron energy spectrum for $r_c=26 \, \rm \mu m$ at creation time (black). The total charge is $1.3\, \rm pC$, and $N_{\rm max} = 6 \, \rm pC/GeV$. Positron spectrum at $z=340 \, \rm \mu m$. The total charge is $17 \, \rm fC$, and $N_{\rm max} = 12 \, \rm fC/GeV$.
    The normalisation facilitates the scaling of the result to higher electron beam charge.
    }
    \label{fig:figure_4}
\end{figure}

The charge content and energy spectrum of the positron beam are summarised in Fig.~\ref{fig:figure_4}.
The number of positrons is shown as a function of the laser propagation distance within the channel.
Most positrons are created in the first $\sim 100 \, \rm \mu m$ of propagation. 
From $100$ to $220 \, \rm \mu m$, the electrons are being injected, forming a tenuous beam of increasing density.
The channel fields are still defocusing positrons during this time, thus explaining the large decrease of $N_{e+}$.
From $220$ to $420 \, \rm \mu m$, the number of positrons is constant since the electron beam loading is large enough to create the field structure guiding positrons.
Beyond $420 \, \rm \mu m$, the guiding structure cannot be maintained as the laser gets depleted ($\gtrsim 30 \%$ energy loss).
The acceleration distance in experiments should therefore be limited below $400 \, \rm \mu m$, to avoid reduction of the injected charge.

The plasma channel size influences the injection process.
For the channel with a small radius ($13 \, \rm \mu m$), ions are loaded at the channel center on top of the positrons.
This motion reduces the negative charge separation field and defocuses positrons.
However, for larger channel widths, the positrons can be guided efficiently as the ions lag behind the positrons.
Please note that the plasma is assumed to be a fully ionized nitrogen gas, with a charge-to-mass ratio of 1/2. If one were to use a lighter species (e.g. protons, with a charge-to-mass ratio of 1), the time scale for acceleration would be smaller, and a wider channel would be required to form a similar guiding structure.
For the case of a nitrogen plasma with a radius $r_c=26 \, \rm \mu m$, the total charge of the positron beam after $340 \, \rm \mu m$ of propagation is $17 \, \rm fC$. Its energy spectrum shown as an inset in Fig.~\ref{fig:figure_4}
evidences positron acceleration from 0.22 GeV at creation time, up to 1.3 GeV after $340 \, \rm \mu m$ of laser propagation.
Higher energies $\sim 5$-$10 \, \rm GeV$ could be reached in a future work, based on scaling laws for electron DLA with multi-PW lasers~\cite{NJPJirka2020}.

\section{Discussion and conclusion}

We now compare our findings with other schemes one can find in the literature.
Post-acceleration setups provide a higher charge and beam quality~\cite{NatCorde2015,NCGessner2016,SRDoche2017,PRLLindstrom2018,PRLJain2015,PRLVieira2014,PPCFLi2019,PRRHue2021,PoPChiou1995,PRLSchroeder1999,PRLWang2008,PRSTABKimura2011,SRYi2014,PRABDiederichs2019,PRLSilva2021,PREReichwein2022}, but they are not compact as they require a kilometer-long accelerator. 
The strength of our method is that positrons are both created and accelerated to a GeV level in a single, sub-millimeter stage.
Other setups that consider multi-PW lasers, either focus only on pair creation without acceleration~\cite{PRLSokolov2010,PRABLobet2017,PRABlackburn2017,arXivStreeter2022} or use 2D geometry which does not allow for reliable quantitative predictions for  the number and energy of the accelerated positrons~\cite{PRLChen2010, PPCFYan2017,SRVranic2018,CPXu2020,CPHe2021}.
For the first time, our Quasi-3D approach allows to obtain quantitative estimates for a compact single-stage positron creation and acceleration scheme.

In conclusion, our work opens the possibility for direct laser acceleration of positrons in a plasma.
We demonstrate that it is possible to create, inject and accelerate positrons in a single-stage experiment by laser-electron scattering at 90 degrees.
We develop an analytical model to estimate the number of created positrons in this geometry and determine with simulations that about  $4 \%$ of these particles are injected and accelerated.
The essential feature for positron guiding is the self-injected electron beam in the channel center ($\sim 800 \, \rm nC$).
It enables positron acceleration to multi-GeV energies in less than a millimeter of laser propagation.

The presented positron acceleration scheme can be realized with the next generation of laser facilities that will reach 75-200 PW~\cite{HPLSEBromage2019,BookMourou2011,OLShao2020,EPJSTBashinov2014,OCMourou2012}.
A minimum power of  $\sim 80 \, \rm PW$, is  required to drive a copious pair production and an efficient positron injection.
Increasing the power further would enhance the number of injected positrons.
The incident electron beam considered (10 pC at 2 GeV) can already be produced experimentally~\cite{PRLGonsalves2019,NCWang2013,PRLKim2013,SRKim2017}.
For the preformed plasma, we suggest to use a commercially available, dense ($10^{21} \, \rm /cm^3$) and short ($400 \, \rm \mu m$) plasma jet~\cite{RSISylla2012}, and to control the channel radius as in Ref~\cite{PoPPieronek2020}.
The femtosecond and micrometer synchronization was already achieved~\cite{PoderPRX2018} and could be less restrictive using larger-scale, laser-driven DLA electron beams~\cite{PRLGahn1999,PRLJiang2016,PPCFShaw2018,NJPWillingale2018,MRERosmej2021,NJPHussein2021, NCGunther2022}.
Our setup generates a broadband beam of GeV-class electrons, positrons and gamma-rays, which opens a new avenue of opportunities for QED cascade seeding or study of propagation of a fireball jet within a plasma in 
a laboratory setting.

\begin{acknowledgments}
  The authors acknowledge fruitful discussions with Prof. L. O. Silva and Mr. Ó. Amaro.
  This work was supported by the European Research Council (ERC-2015-AdG Grant
  695088).
  We acknowledge the support of the Portuguese Science Foundation (FCT) Grant No. CEECIND/01906/2018 and PTDC/FIS-PLA/3800/2021.
  We acknowledge PRACE for awarding access to MareNostrum based in the Barcelona Supercomputing Centre.
\end{acknowledgments}

%

\pagebreak
\widetext
\begin{center}
\textbf{\large Supplementary Material: Creation and direct laser acceleration of positrons in a single stage}
\end{center}
\setcounter{equation}{0}
\setcounter{figure}{0}
\setcounter{table}{0}
\setcounter{section}{0}
\makeatletter

\centerline{Bertrand Martinez, Bernardo Barbosa, Marija Vranic}
\centerline{\textit{Golp/Instituto de Plasma e Fus\~{a}o Nuclear, Instituto Superior Técnico}}
\centerline{\textit{Universidade de Lisboa, 1049-001 Lisbon, Portugal}}

\section{Modelling geometry}

This work relies on Quasi-3D modelling of Osiris~\cite{JCPDavidson2015}.
In theory, quasi-3D modeling could recover a full Cartesian 3D description of the plasma fields if all modes $m\geq0$ would be included (the Fourier decomposition makes for a complete basis).
In practice, the first two modes are sufficient to describe the dynamics of the self-generated plasma fields ($m=0$) as well as the linearly polarised laser field ($m=1$).
In the quasi-3D algorithm, particles are pushed in 3D Cartesian geometry, but each Fourier mode of the electromagnetic field is represented on a two-dimensional lattice.
The computational cost of such a simulation is thus comparable to the 2D simulation requirements and also retains some important information about the 3D dynamics.

It is worth stressing that quasi-3D modelling goes beyond a 2D approach and retains some important information about the 3D dynamics.
For example, the electromagnetic fields in 2D decay slower with distance than in 3D, which makes it difficult to make reliable quantitative predictions regarding nonlinear acceleration processes. 
Quasi-3D does not have this problem and it can thus model correctly the laser focusing and defocusing, even in nonlinear interaction of light with plasmas. 

Simulations in this Letter were conducted with the quasi-3D version of the Particle-In-Cell (PIC) code Osiris~\cite{IPFonseca2002,JCPDavidson2015}.
To this end, this version was recently merged with the QED-PIC algorithm to account for gamma-ray emission via non linear inverse Compton scattering as well as Breit-Wheeler pair production in the local constant field approximation.

\section{Simulation parameters} \label{sec:appendix2}

The laser pulse propagates along the axis $z$ (the $\parallel$ direction) and is linearly polarised in direction $x$ (the $\perp$ direction).
The laser field has a temporal duration with a Full Width at Half Maximum (FWHM) of $\tau_L = 150 \, \rm fs$ and waist of $w_L = 3.2 \,\rm \mu m$.
For the  $\lambda_L=1 \, \rm \mu m $ laser wavelength we consider here, we choose a peak intensity of $I_L = 5 \times 10^{23} \, \rm W/cm^2$.

The preformed plasma channel is initialized as a fully ionized, neutral nitrogen plasma of temperature $2 \, \rm eV$.
Its radial profile has a parabolic shape and we considered a linear density gradient of length $L=6 \, \rm \mu m$ at the longitudinal boundary 
 $ n(z,r)/n_0 = \left[10^{-3} + (4-10^{-3}) (r/r_{c})^2\right]\times \left(z/L\right)$ with $r_c = 26 \, \rm \mu m$ the channel radius.
In practice, this means the electron density on the central axis is $10^{-3} n_0 = 10^{18} \, \rm /cm^3 $ and increases up to $4 n_0$ on the channel walls, at $r = r_c$.
The quantity $n_0=m\epsilon_0\omega_L^2/e^2 \sim 10^{21} \, \rm /cm^3$ is the critical density for the laser of frequency $\omega_L$, with $m$ the electron mass, $\epsilon_0$ the vacuum permittivity and $e$ the electron charge.

The electron beam propagating along the x axis cannot be represented in the quasi-3D geometry using low order modes as it carries a transverse current and density distribution.
For this reason, we initialise only the gamma-rays that do not carry any current.
The spatial distribution of photons in the longitudinal direction (z) is Gaussian with a FWHM of $3 \, \rm \mu m$, comparable to a laser-driven electron beam.
The photon radial distribution is uniform for computational purposes, which adds additional photons outside the interaction region.
As the pairs are created only in focus, these additional photons do not significantly affect the number of created positrons.
We derived a conservative lower bound for the photon number density as follows.
We assume 10 pC of charge in the incident beam ($6\times 10^7$ electrons) and that only $10 \%$ of them will radiate due to the geometry of interaction (this estimate is based on previous results from full-3D simulations).
As a result there are, at least, $6\times 10^6$ gamma-rays created in the interaction region that has a volume of $10 \times 1 \times 1 \, \rm \mu m^3$ (same as a LWFA electron beam).
This amount to a gamma-ray density of $2\times 10^{17} \, \rm /cm^3$.
We then take $10^{17} \, \rm /cm^3$ as a conservative lower bound estimate.

\begin{figure}
    \centering
    \includegraphics[width=0.45\textwidth]{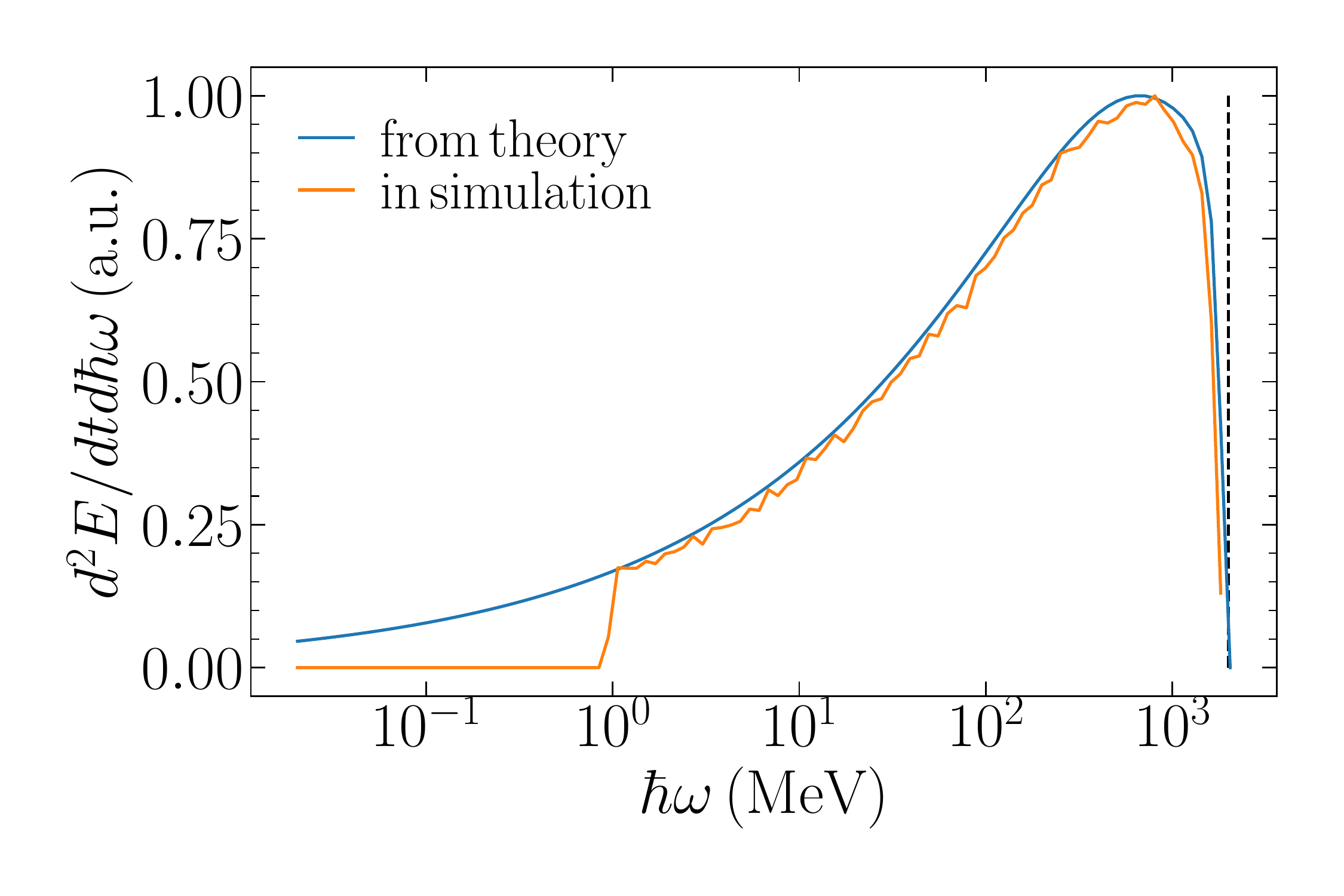}
    \caption{
    Energy spectrum of photons initialised in our simulations, and as provided in theory~\cite{RMPErber1966}. The incident electron energy is 2 GeV and its quantum parameter is $~3$.}
    \label{fig:figure_5}
\end{figure}

The energy distribution of gamma-rays is initialised with a synchrotron spectrum illustrated in Fig.~\ref{fig:figure_5}.
The analytical expression comes from Ref~\cite{RMPErber1966}.
Photons with an energy below 2 MeV cannot create electron-positron pairs and are thus not initialised nor propagated.

For the spatial domain, we use a moving window (at light velocity), following the pulse propagation.
The spatial steps are $\Delta z=\Delta r= 16 \, \rm nm$, and the simulation domain extends over $204.8 \times 96 \, \rm \mu m^2$.
Simulations run for a duration $1.46 \, \rm ps$ with a time step $26.6 \, \rm as$, which corresponds to a propagation distance of $438 \, \rm \mu m$.
Initially, we use $16/16/128$ particles per cell respectively for electrons, ions, and gamma-rays.
Breit-Wheeler pair production and radiation reaction in the quantum regime are accounted for with a MC module.

\section{Pair production at an arbitrary angle}
    
The function $P_\pm$ in Eq.~(1) is derived using the method applied in previous works~\cite{PRABlackburn2017,NJPMercuri2021,NJPAmaro2021}.
It is a generalisation for any angle of incidence between the laser pulse and the gamma-ray.

Let us consider a plane wave propagating along axis $z$ and linearly polarised along axis $x$.
Its wavelength is $ \lambda_L$ and we denote by $\mathbf{k}_L$ its wave vector.
Its angular frequency is denoted by $\omega_L$.
The field of this plane propagating wave is thus simply given by the expression $\mathbf{E}(z,t) = E_L \sin( \omega_L t - k_L z) \mathbf{x}$, with $E_L$ the peak amplitude of the field.

Let us consider also a single gamma ray propagating with an angle $\psi \in (0,180^\circ)$ with respect to the plane wave propagation axis z.
It has a normalised energy $\gamma_\gamma = \hbar \omega / mc^2$, where $\hbar$ is the reduced Planck constant, $\omega$ is the photon angular frequency, $m$ the electron mass and $c$ the light velocity.
We define its wave vector as $\mathbf{k}_\gamma = \gamma_\gamma( \sin(\psi) \mathbf{x} + \cos(\psi) \mathbf{z})$
The angle of interaction between the laser pulse and the gamma ray is defined as $ \cos \psi = \mathbf{k}_L \cdot \mathbf{k}_\gamma / |\mathbf{k}_L|| \mathbf{k}_\gamma| $.
The quantum parameter of the photon is denoted $\chi_\gamma$.

The pair production rate of a photon is given by Ritus~\cite{JSLRRitus1985} and denoted $dN_{BW}/dtd\chi_\gamma$.
In a similar manner as in an earlier study~\cite{NJPMercuri2021}, we introduce the convenient function $b_0$ defined as $dN_{BW}/dtd\chi_\gamma= W_0 b_0(\chi_\gamma)/\gamma_\gamma$, or, in other words,
\begin{equation}
        b_0(\chi_\gamma) = \frac{\sqrt{3}}{2\pi} \int_0^1 \frac{d\xi}{\xi(1-\xi)} \Bigg[  \frac{2}{3\chi_\gamma} \frac{1-2\xi}{1-\xi} K_{5/3}(\mu) + K_{2/3}(\mu) \Bigg]
\end{equation}
where $\mu=2/\left[3\chi_\gamma\xi(1-\xi)\right]$, and the constant $W_0$ is defined as $W_0 = 2\alpha mc^2/(3\hbar)$.

Using the definition of the quantum parameter of the photon, its temporal evolution in the plane wave can be derived
\begin{equation}
    \chi_\gamma(t) = \frac{2}{c_\psi} \,  \frac{\gamma_\gamma E_L }{ E_S } \,  \left\lvert\sin\left( \frac{2 \omega_L t }{ c_\psi} \right)\right\rvert
\end{equation}
where we introduce the geometrical factor $c_\psi = 2 / (1-\cos \psi)$, characterizing the angle of incidence between the gamma-ray and the plane wave.
The function $\chi_\gamma(t)$ reaches local maxima denoted $\chi_\gamma^{(m)} = 2 \gamma_\gamma E_L^{(m)} / ( c_\psi E_S ) $ with a period of $ \omega_L T = \pi c_\psi/2 $, where $E_L^{(m)}$ denotes the peak field amplitude of the $m$-th cycle of the pulse.
For the $m$-th cycle, with $ m \geq 0$, this maximum is reached for $\omega_L t_m = \pi c_\psi/4 + m\pi $ and the probability of pair production for a single gamma ray can be easily expressed on the period $T$, if we center the integral on the maximum $t_m$
\begin{equation}
    P_{\pm}^{(m)} = 1 - \exp\left( - \int_{t_m-T/2}^{t_m+T/2} W_{\pm} \left(\chi_\gamma(t)\right) \, dt\right)
\end{equation}
We change the variable $\phi=2\omega_L t / c_\psi$ and recast it as
\begin{equation}
    P_{\pm}^{(m)} = 1 - \exp\left( - \, \frac{c_\psi W_0}{2\omega_L \gamma_\gamma} \int_{0}^{\pi} b_0 \left(\chi_\gamma^{(m)} |\sin(\phi)|\right) \, d\phi \right)
\end{equation}
The evaluation of this integral can be done either numerically, or be approximated to a good degree with the saddle point approximation.
Let's denote it $I_0$ such as
\begin{equation}
    I_0 = \int_{0}^{\pi} b_0 \left(\chi_\gamma^{(m)} \left \lvert \sin(\phi) \right \rvert\right) \, d\phi
\end{equation}
The detailed calculation for $ \psi=180^\circ $ is done in~\cite{NJPMercuri2021} and we generalised it for any $ \psi \in (0,180^\circ) $.
This relies on the fact that the main contribution of the integral comes from the maximum of the integrand, reached for $\phi_m = \pi / 2$.
Following the saddle point method, we can approximate the integrand close to this maximum by
\begin{equation}
    b_0\left(\chi_\gamma^{(m)}|\sin \phi|\right) \simeq b_0\left(\chi_\gamma^{(m)}\right) \exp\left(-\frac{(\phi-\phi_m)^2}{2s^2\left(\chi_\gamma^{(m)}\right)}\right)
\end{equation}
where $s\left(\chi_\gamma^{(m)}\right)$ is chosen so that the exact and approximated integrands have the same second derivative at $\phi=\phi_m$.
This condition provides $s(\chi) = c_\psi \sqrt{b_0(\chi)/\chi b_0'(\chi)}$, where $b_0'$ is the derivative of the function $b_0$ with respect to $\chi$.
The integral can now be performed, using the approximated expression for the integrand and leads to
\begin{equation}
    I_0 = \pi b_0\left(\chi_\gamma^{(m)}\right) \min\Big[F\left(\chi_\gamma^{(m)}\right),f(\psi)\Big]
\end{equation}
with  the function $F$ given by the expression
\begin{equation} \label{eq:Fschi}
        F(\chi) = \frac{s(\chi)}{\sqrt{2\pi}} \Bigg[  \erf\left(c_\psi\frac{\pi}{2}\frac{1}{\sqrt{2}s(\chi)}\right) + \erf\left(\left[\pi-c_\psi\frac{\pi}{2}\right]\frac{1}{\sqrt{2}s(\chi)}\right) \Bigg]
\end{equation}
where $\erf$ denotes the error function.
The function $f_\psi$ is introduced~\cite{NJPMercuri2021} to bring a correction in the regime $\chi_\gamma \gg 1$ and can be generalised for any angle $\psi$ by the expression $ \pi f(\psi) = \int_0^\pi\sin^{2/3}(\phi/c_\psi) \, d\phi$.
The probability for the full pulse, including all its cycles obtained is a purely analytical formula and it depends on the following variables: $\omega_L$ the laser angular frequency and $\chi_\gamma^{(m)}$ for $m\geq 0$, the peak quantum parameter of the photon in the $m$-th cycle of the pulse
\begin{equation} \label{equation:2}
    P_\pm = \sum_{m\geq 0} P_{\pm}^{(m)} \text{ with } P_{\pm}^{(m)} = 1 - e^{-\frac{1}{\omega_L}W_\pm\left(\chi_\gamma^{(m)}\right) F\left(\chi_\gamma^{(m)}\right)}
\end{equation}
This probability is valid for one photon interacting with a plane wave. It is used in Eq~(1) of the article in order to derive the number of positrons expected for a focused pulse interacting with a photon beam, which has a Gaussian shape in the z direction ($3 \, \rm \mu m$ FHWM) and is uniform in the radial direction.

\section{Maximizing the number of positrons available for acceleration}

The deflection of positrons created along the laser propagation direction depends on the number of low energy positrons generated at the interaction point~\cite{SRVranic2018}.
This number is intuitively maximized for an incident electron beam of low energy $\mathcal{E}_{e-}$.
However, this energy $\mathcal{E}_{e-}$ cannot be too low, otherwise no pairs are created by the Breit-Wheeler process.
There is therefore an optimal value of $\mathcal{E}_{e-}$ that can be evaluated.

We consider that the time interval $\Delta t$ that the laser spends interacting with the electron beam is
$ \Delta t = w_L/c \simeq 13 \, \rm fs$, corresponding to a light crossing time in the laser pulse at a $90^\circ$ angle of incidence, with $w_L=3.8 \, \rm \mu m$ denoting the transverse FWHM of the laser field.
As for the electron beam, we consider that it has a total charge of $10 \, \rm pC$.
We assume that photon and pair production occur uniformly during the time interval $\Delta t$, such that the number of produced particles
of differential rates $dN_{BW}/dtd\gamma_+$ is given by $\Delta t \times dN_{BW}/dtd\gamma_+$.

\begin{figure}
    \centering
    \includegraphics[width=0.45\textwidth]{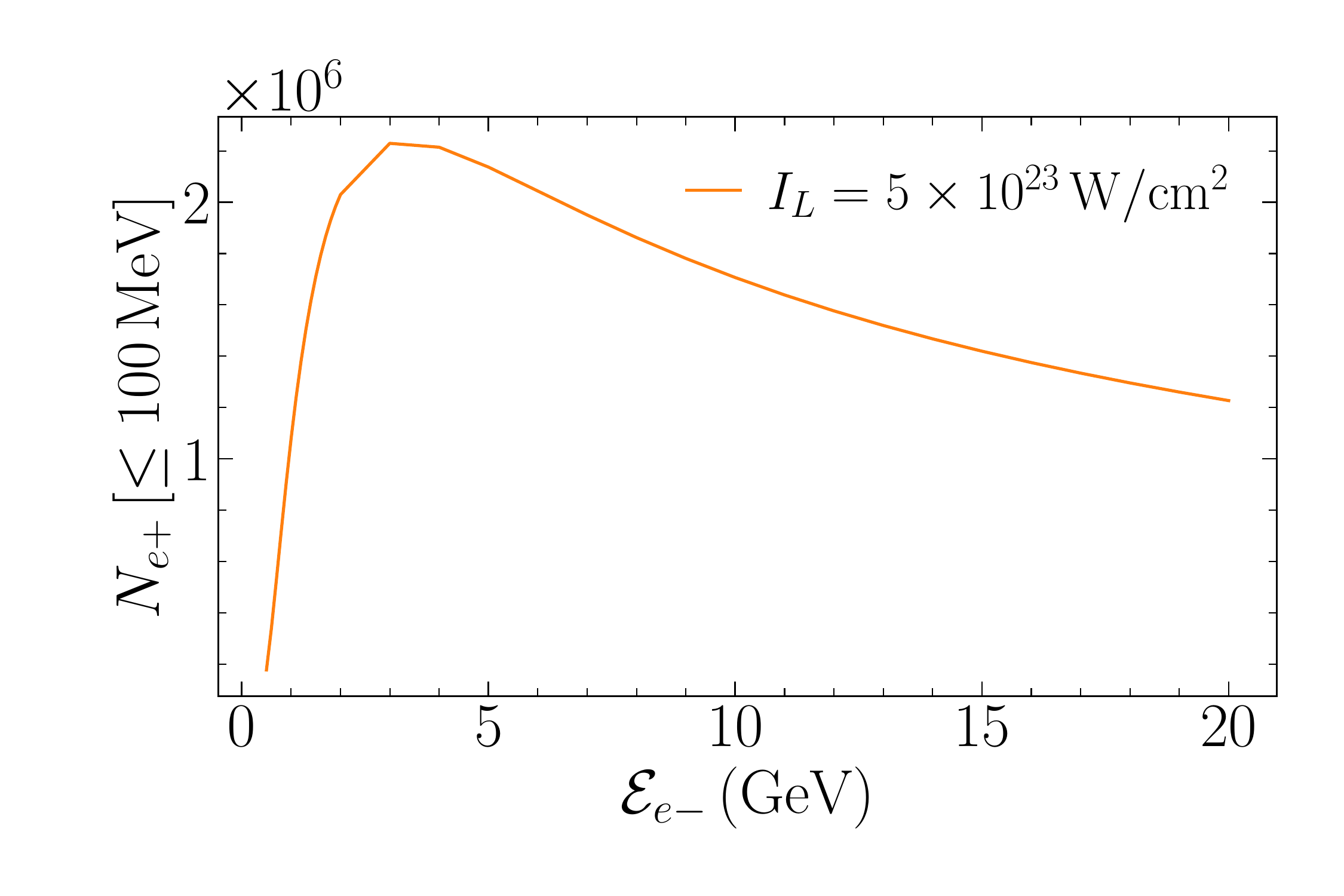}
    \caption{
    Number of low energy positrons ($\leq 100 \, \rm MeV$) generated during the interaction of a plane wave of normalized amplitude $I_L=5\times 10^{23} \, \rm W/cm^2$ with a mono energetic electron beam of energy $\mathcal{E}_{e-}$.
    }
    \label{fig:figure_6}
\end{figure}

For a waterbag distribution of photons with a normalized energy $\gamma_\gamma$, the positron energy spectra is derived by multiplying the number of pairs produced by the pair production rate
\begin{equation*}
    \frac{dN_w}{d\gamma_+} = \left(\Delta t\right)^2 \frac{d^2N_{BW}}{dtd\gamma_+} \frac{dN_{BW}}{dt}
\end{equation*}
The positron spectra for a synchrotron energy distribution is deduced by weighting this spectra by the nonlinear Inverse Compton scattering differential cross-section $dN_{IC}/dtd\gamma_\gamma$~\cite{RMPErber1966}
\begin{equation*}
    \frac{dN_{IC}}{d\gamma_+} = \Delta t \int d \gamma_\gamma \, \frac{d^2N_{IC}}{dtd\gamma_\gamma} \frac{dN_w}{d\gamma_+}
\end{equation*}
The above equation provides the positron energy distribution function emitted by a mono energetic electron beam.
In order to quantify the number of low energy positrons, we integrate it numerically from $0$ to $100 \, \rm MeV$, using the routines from the Boost library for C++~\cite{BookSchling2011}.
The calculation is performed for a laser intensity $I_L=5\times 10^{23} \, \rm W/cm^2$ and for a large range of electron beam energies $ \mathcal{E}_{e-} = 0.5  \rightarrow 20 \, \rm GeV $.
The result is displayed in Fig.~\ref{fig:figure_6} and evidences the presence of an optimum electron beam energy of $3 \, \rm GeV$.
This optimum is the value ensuring that a maximum of positrons are deflected in the laser propagation direction.

\end{document}